\documentclass[12pt]{article}
\usepackage[margin=1.5in]{geometry}

\usepackage[utf8]{inputenc} 
\usepackage[T1]{fontenc}    
\usepackage{hyperref}       
\usepackage{url}            
\usepackage{booktabs}       
\usepackage{amsfonts}       
\usepackage{nicefrac}       
\usepackage{microtype}      
\usepackage{lipsum}
\usepackage{fancyhdr}       
\usepackage{graphics}
\usepackage{graphicx}       

\usepackage{pifont}
\usepackage{xcolor}

\title{\#Secim2023: First Public Dataset for Studying Turkish General Election}

\author{
  Ali Najafi \and Nihat Mugurtay \and Ege Demirci \and Serhat Demirkiran \and Huseyin Alper Karadeniz \and Onur Varol\footnote{Corresponding author: \texttt{onur.varol@sabanciuniv.edu}}
}

\date{%
    Sabanci University \\ 
    \today
}

\begin{document}
\maketitle

\begin{abstract}

In the context of Turkey's upcoming parliamentary and presidential elections (``seçim'' in Turkish), social media is playing an important role in shaping public debate. The increasing engagement of citizens on social media platforms has led to the growing use of social media by political actors. It is of utmost importance to capture the upcoming Turkish elections, as social media is becoming an essential component of election propaganda, political debates, smear campaigns, and election manipulation by domestic and international actors. 
We provide a comprehensive dataset for social media researchers to study the upcoming election, develop tools to prevent online manipulation, and gather novel information to inform the public. We are committed to continually improving the data collection and updating it regularly leading up to the election. Using the \texttt{Secim2023} dataset, researchers can examine the social and communication networks between political actors, track current trends, and investigate emerging threats to election integrity.
\\
Our dataset is available at: \href{https://github.com/ViralLab/Secim2023_Dataset}{github.com/ViralLab/Secim2023\_Dataset}

\end{abstract}

\section*{Introduction}

In recent years, political debates are increasingly taking place on social media platforms, and their impact on political behavior has been heavily discussed in the literature~\cite{jungherr2014twitter,metaxas2012social,morgan2018fake,bilal2019predicting}. 
This has led to a variety of research in political science, including election forecasting, public opinion, political network detection, and election manipulation~\cite{digrazia2013more, jahanbakhsh2014predictive, deb2019perils, anstead2015social, o2010tweets,jaidka2019predicting, wang2015forecasting, ratkiewicz2011detecting, majo2021role}.  
As people use social media extensively, various automated approaches are used by politicians, political parties, and voters to garner support and influence a country's political agenda or manipulating online discourse by spreading fake news and misinformation~\cite{peel2013coalition,yang2019arming,shao2018spread,cresci2020decade,morgan2018fake,faris2017partisanship,vosoughi2018spread,lazer2018science}. 
Similarly, citizens' exposure to social media contributes to the spread of conspiracy theories, which vary by ideological affiliation~\cite{ccalicskan2022behind}. 
In some countries, governments also work with Twitter to censor content and limit the visibility of that content to their citizens~\cite{tanash2015known,varol2016spatiotemporal}.
In this regard, Twitter is an influential social media platform that can influence citizens' political engagement, and studying online trends and offline results in elections has become increasingly interesting in recent years\cite{fujiwara2021effect}.

At a time when Turkey is preparing for the upcoming presidential and parliamentary elections, social media and digital propaganda are becoming increasingly important. The number of Turkish social media users has increased at an unprecedented rate, which can reach a point of addiction among young people \cite{kirik2015quantitative}. The majority of these young users will be new voters in the upcoming Turkish elections. Using our novel Twitter dataset, we aim to reveal the political trends during pre-election and the campaign of the next Turkish elections. Despite the growing literature on social media data and elections, there is still a lack of empirical evidence explaining the key online dynamics during the upcoming elections in Turkey.

In this paper, we present our methodology for collecting a comprehensive social media dataset to study the upcoming election.
We operationalize this raw dataset to capture daily user activity, volume of tweets, city-level trending topics, network activity, and ego-centric networks. In addition, we also provide an empirical analysis of bot activity observed on politicians' social networks.

\section*{Background on Upcoming Elections}

Turkish politics underwent unprecedented change through referendums, parliamentary/presidential and local elections. Access to alternative sources of information is an essential component of a functioning democracy during the process of free and fair elections~\cite{dahl2005political}. The use of social media has become a significant aspect of public debate on social and political issues, thwarted by the gradual control of mainstream media by government-affiliated corporations~\cite{bulut2017mediatized}. People are finding new ways to connect and gather information through the use of various modalities of online platforms~\cite{hoyng2017conspiratorial}. In Turkey, this extensive use of social media occurs within the context of debates about popular protests, regime oscillations, polarization, populism, press-party parallelism, and social media manipulation~\cite{esen2016rising,metin2022tweeting,toros2022social,bulut2017mediatized,toros2015negative,yesil2021social,varol2014evolution,yaman2014gezi,saka2016siyasi, irak2016close,ogan2017gained}.

The upcoming elections are of paramount importance to all political parties. Following the adoption of the 2018 ``Alliance'' article in the electoral law\footnote{Law Clause 12/A. \url{https://www.mevzuat.gov.tr/mevzuat?MevzuatNo=2839&MevzuatTur=1&MevzuatTertip=5}} by the Turkish Parliament, two alliances emerged with strong electoral and legislative implications~\cite{moral2021story}. The new election legislation stipulates that presidential and parliamentary elections take place on concurrently for every five years\footnote{Law Clause 3 of new \#7140 amendment: https://www5.tbmm.gov.tr/kanunlar/k7140.html}. While Parliamentary elections take place using the conventional proportional party-list system, citizens elect the president using a two-round majority method. Moreover, there is a 7 percent threshold for parliamentary elections \footnote{It was 10 ten percent between 1982 and 2022}. This leads to a winner-take-all scenario, in case a political alliance win both presidential and parliamentary elections. Recently, ruling People's Alliance does not have power to change constitution unilaterally in the parliament. The Justice and Development Party (AKP) and the Nationalist Movement Party (MHP) will both benefit greatly from winning next general elections as this will allow them to remain in power and preserve their alliance.\footnote{Nationalist Great Unity Party (BBP) is also a small member of alliance} 
The main opposition, known as National Allegiance, consists of four main political parties including Republican People's Party (CHP), Good Party (İYİP), Felicity Party (SP) and Democrat Party (DP). Two political parties detached from the ruling party AKP, Democracy and Progress Party (DEVA) and Future Party (GP) also support Nation Alliance.\footnote{Turkish abbreviations: \textbf{AKP}:Adalet ve Kalkınma Partisi, \textbf{MHP}: Milliyetçi Hareket Partisi, \textbf{CHP}: Cumhuriyet Halk Partisi, \textbf{İYİP}: İyi Parti, \textbf{SP}: Saadet Partisi, \textbf{DP}: Demokrat Parti, \textbf{DEVA}: Demokrasi ve Atılım Partisi, \textbf{GP}: Gelecek Partisi, \textbf{BBP}: Büyük Birlik Partisi, \textbf{HDP}: Halkların Demokratik Partisi, \textbf{National Alliance}: Millet İttifaki, \textbf{People's Alliance}: Cumhur İttifaki} The third group is represented by People's Democratic Party (HDP), which is composed of mostly Kurdish electorate and various leftist political parties. The next Turkish parliamentary and presidential elections will occur at a time when economic and political problems and polarization are at their peak~\cite{esen2022}. It is during this period that all actors are trying to influence public opinion via social media, either by attracting citizens or provoking negative campaigns against their opponents. In general, electoral campaigns intersects electorate's social media exposure, by which ideological and political groups buy the propaganda, information and conspiracy.   

This electoral process is supported by the extensive use of social media by the population. In this paper, we examine the upcoming Turkish elections by highlighting the key social media trends for the pre-election and campaigning processes. We adopt Norris \textit{et al}(2019)'s  phases of election, which include the pre-election, election campaign, election day, and post-election periods~\cite{norris2019electoral}. Each period features different modes of political dynamics. For example, the first two phases include negative campaign strategies, election promises and individual criticism towards candidates, debates over election laws, media portrayals of each party, and campaign financing. Therefore, the timing of our study is also suitable for describing the online political behavior of citizens and politicians, i.e., the dynamics of citizens' political behavior in social media can be best captured during this phase.

Previously, several researchers have looked at how to use Twitter data for political analysis \cite{savacs2021big,gokcce2014twitter,bulut2017mediatized,ozturk2018sentiment,polat2014twitter,bozdag2014does,ozaydin2021fragmentation}. The techniques used in these studies include a wide range of computational tools, including network, sentiment, and content analysis. Our contribution to the literature is based on three key pillars of big-data analysis. As a first step, we provide a novel dataset on Twitter and other online sources that is essential for understanding the main dynamics of the upcoming Turkish parliamentary and presidential elections. Using Twitter Developer APIs, our dataset captures both user data and influential political figures in Turkey. Second, we provide an initial analysis of social networks, daily changes in political figures' followers, trending topics, the number of daily tweets from individual users, and party membership data by time.
Thus, our raw data combine empirical and technical aspects of computational tools with political science. Finally, we discuss the implications of our study for further operationalization of the dataset. We contribute to the literature not only by providing a structured dataset, but also by setting a research agenda for how it can be used to describe existing trends before and during an election.

\section*{Data Collection}

\begin{figure}[t!]
    \centering
    \includegraphics[width=\linewidth]{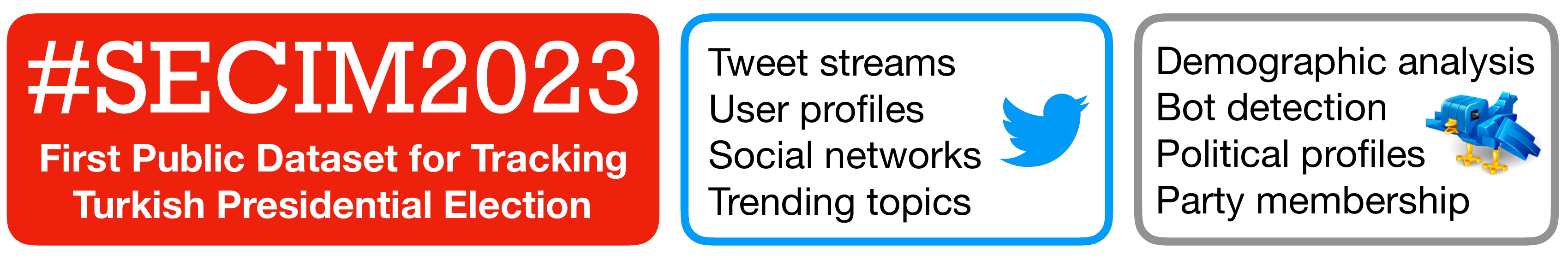}
    \caption{\textbf{Summary of \texttt{\#Secim2023} dataset.}}
    \label{fig:schematic}
\end{figure}

In analyzing conversations about elections and political debates, we rely on predetermined keywords and users. Using available resources, we aim to capture a holistic picture of Turkish elections on Twitter.  
Data collection in this project uses Twitter API versions v1.1 and v2.0. Our team has access to the standard developer API and elevated access via the Academic API. We use the \texttt{tweepy} library for Python to systematically access the Twitter API. We also developed custom web scrapers to download additional information such as party membership statistics. Schematic in Fig~\ref{fig:schematic} summarizes the different data sources.

\textbf{Trending topics}:
Twitter provides ''trending topics'' on the platform to share important conversations. These trending topics can be hashtags or phrases, and are available at the city and country level as well as globally. We use Twitter's Trending Topics API to collect hashtags and phrases for 12 cities available in Turkey and country-level trends. We collect trends every 10 minutes to systematically track changes in conversations.

\textbf{Twitter annotation streams}:
Twitter Academic API introduced tweet annotations\footnote{https://developer.twitter.com/en/docs/twitter-api/annotations/overview} where named entities comprised of people, places, products, and organizations are automatically detected. Twitter use these entities to link with various topics including politics. We selected those context annotations about politics (\texttt{context:35.*}, \texttt{context:38.*}, \texttt{context:88.*}) and filtered the ones that are written in Turkish. Since the volume of activity is quite significant and the Academic API limits us with 10 million tweets per month, we collect random sample of 25\% of the retweeted content and keep all original tweets, quotes, and replies. 

\textbf{Streaming API}: Although the entity annotation feature is useful for data collection on a particular topic, our initial analysis shows that Twitter's entity detection system systematically biased towards the members of the current government. Since we want to capture all political discussions in our dataset, we create our own keywords and users lists for collecting data from the streaming API. Our collection of political users includes party leaders, mayors of the major cities, and members of the Grand National Assembly from the last two terms. Although this list is comprehensive and currently covers 936 different users, we regularly update our list and also collect recommendations via a public form.\footnote{Account recommendation form: \url{https://forms.gle/aPb44xMqvqXZSRim8}}

\textbf{Social networks and profiles}: The Twitter API can provide social network connections. We have collected both friends and followers of the political accounts. 
The API provides these connected accounts in chronological order, based on when the edge was created 
We then collect profile objects of these users for further analysis. In this dataset, we provide meta-data about account profiles such as \texttt{id}, \texttt{name}, \texttt{screen name}, \texttt{account creation time} and statistics such \texttt{friend}, \texttt{follower}, and \texttt{tweet} counts at the time of data collection.

\textbf{Party membership}: Statistics about party members are shared on the website for ``General Prosecutor Office of the Supreme Court of Appeal.\footnote{\url{https://www.yargitaycb.gov.tr/kategori/117/siyasi-partiler}}'' We developed a scraper for this website to collect statistics about memberships daily. 

\textbf{Model based data enrichment}: We also utilize pre-trained models to infer account-level properties such as bot score and demographics using BotometerLite\cite{yang2020scalable} and m3inference\cite{wang2019demographic} systems and tweet-level measures such as sentiment, topics, and named-entities using pre-trained language model. 

\section*{Data Access}

The dataset is publicly available and continuously maintained on Github at this address: \href{https://github.com/ViralLab/Secim2023_Dataset}{github.com/ViralLab/Secim2023\_Dataset}

We are maintaining this dataset on an ongoing basis, and data collection is still in progress. We plan to update the dataset monthly. If you have suggestions for data collection, please use Github issues and our team will consider them and try to incorporate them into the dataset.

\textbf{Note}: Twitter's Developer Agreement \& Policy limits us sharing the full dataset collected through their API. This limits our ability to share entire information about tweets, but instead we provided tweet IDs that can be rehydrated using Twitter's API. Software packages like  Hydrator\footnote{\url{https://github.com/DocNow/hydrator}} or Twarc\footnote{\url{https://github.com/DocNow/twarc}} can be used to systematically download data. Unfortunately, deleted tweets will not be available when collected from API.

\section*{Data Analysis}

Our dataset is comprehensive and unique to Turkish elections in that it collects trending topics, tracks political accounts, and extracts networks and additional signals from these entities. 

\begin{figure}[t!]
    \centering
    \includegraphics[width=\linewidth]{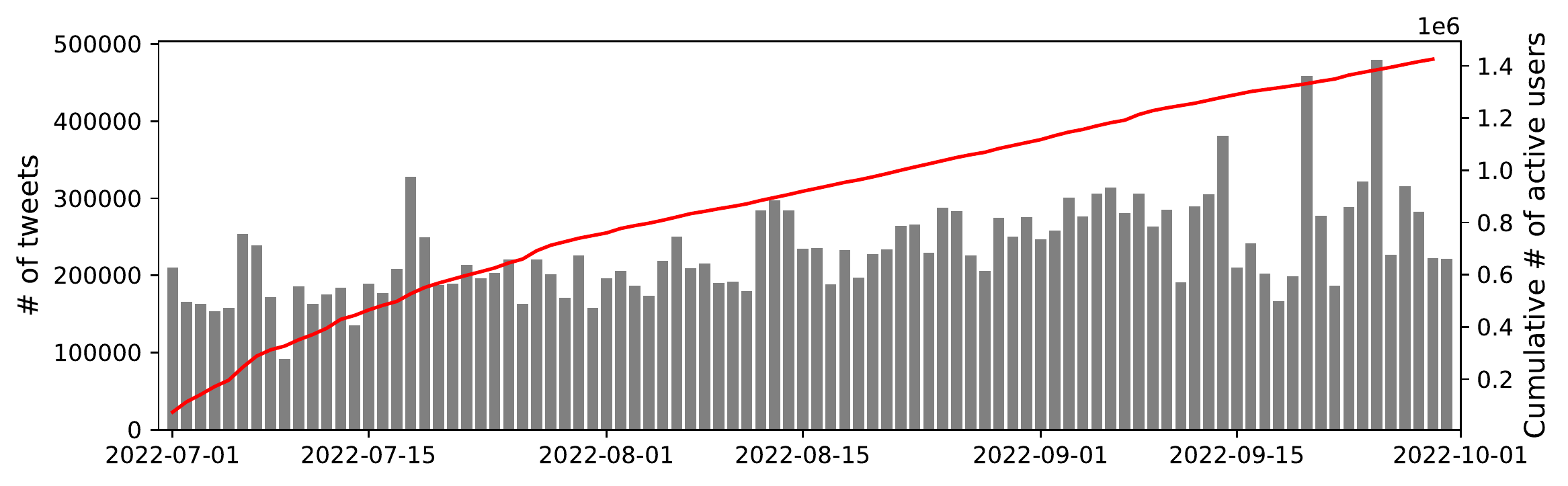}
    \caption{\textbf{Temporal statistics of streaming dataset.} We track number of daily tweets and the unique users sending those tweets.}
    \label{fig:temporal-stats}
\end{figure}

Our data streams provide Turkish tweets that contain political entities or relevant keywords. On average, we collect about 200,000 tweets per day posted by more than 1.4 million accounts in the last three months, as shown in Fig.\ref{fig:temporal-stats}.

\begin{figure}[t!]
    \centering
    \includegraphics[width=\linewidth]{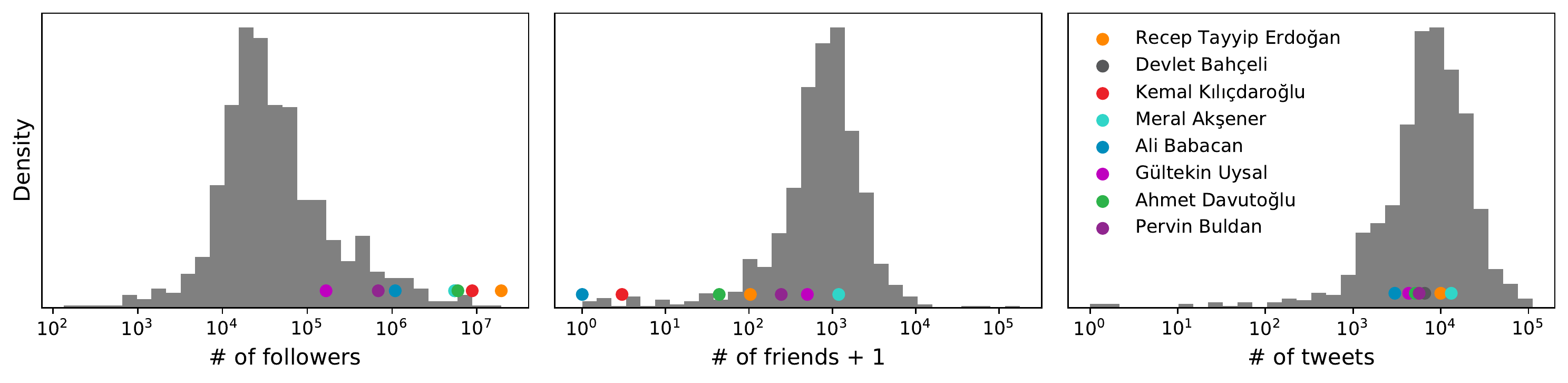}
    \caption{\textbf{Profile statistics of political actors.} Comparing profile metrics such as number of friends, followers, and posts for accounts in our collection.}
    \label{fig:profile-stats}
\end{figure}

We also track the profile statistics of political accounts daily. We can compare their profile characteristics as shown in Fig.\ref{fig:profile-stats}. Exemplar politicians we selected produce average level of content, but are more popular than the rest of the political figures and they are more selective for their friends.

\begin{figure}[t!]
    \centering
    \includegraphics[width=\linewidth]{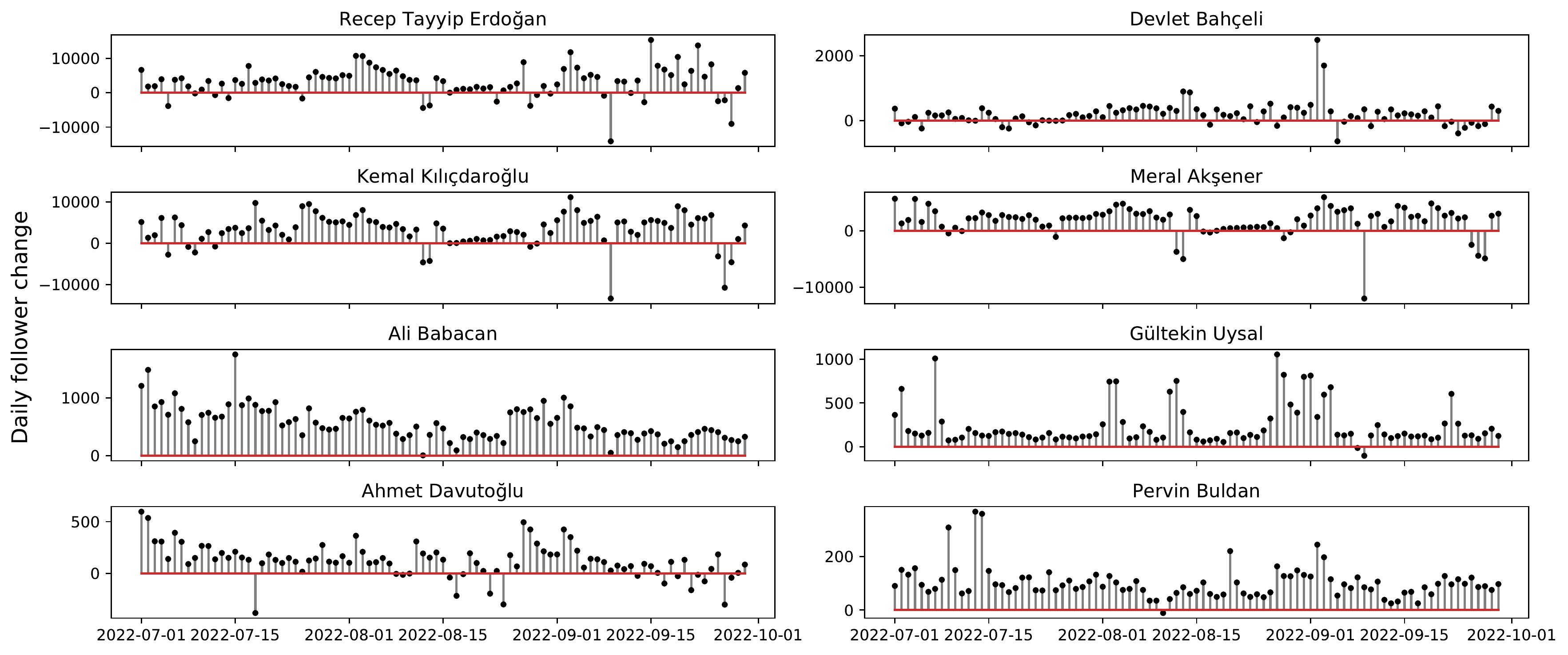}
    \caption{\textbf{Daily follower changes of political figures.} We collect profile information of politicians daily and we can monitor changes in their profile statistics daily.}
    \label{fig:follower-change}
\end{figure}

Since we track the daily changes of politicians profiles, we can look at the daily changes of their followers as a time series. 
In Fig.\ref{fig:follower-change}, we can point the correlated changes of follower counts. For instance Recep Tayyip Erdogan, Kemal Kilicdaroglu, and Merak Aksener lost nearly 10,000 followers on September 9th, 2022. This significant change can be due to deleted accounts or result of an automated coordinated activity.  

\subsection*{Trending topics in Turkey}

Trending topics frequently reflect important events such as sporting events, political debates, or TV in Turkey. They are also shown to be manipulated by means of automation, and a recent study suggests that 47\% of the top 5 daily trends are generated by astroturfing attacks~\cite{elmas2021ephemeral}.

\begin{figure}[t!]
    \centering
    \includegraphics[width=\linewidth]{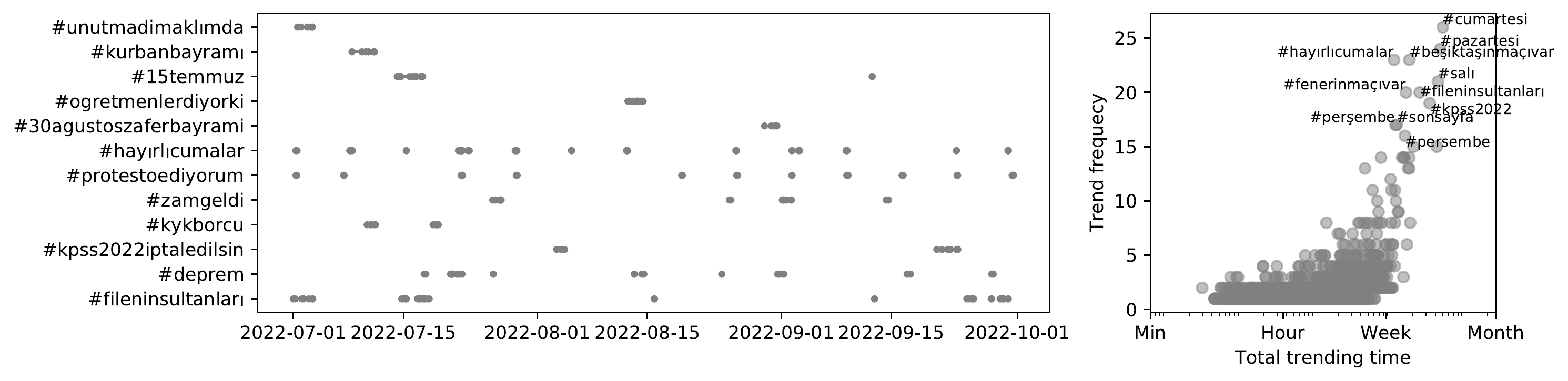}
    \caption{\textbf{Temporal characteristics of trend topics.} Trending topics of a particular location (Turkey in this example) can be visualized as a timeline (left). Frequency and trending time can also be calculated for all trends to identify outliers (right).}
    \label{fig:trend-timelines}
\end{figure}

To capture manipulated trending topics and capture important events, we collect those trends regularly. In Fig.\ref{fig:trend-timelines}(left), we show sample of trending hashtags that appeared in the top-10 list for a significant period. Some of these hashtags point important days such as \texttt{\#unutmadimaklimda} to commemorate Sivas massacre happened in July, 2nd 1993 or \texttt{\#30agustoszaferbayrami} for celebrating Victory day. 

We also studied frequency and duration of trending hashtags in Fig.\ref{fig:trend-timelines}(right). Hashtags for days of the week or football games repeat regularly, while most hashtags stay in the trending list for less than a day.

Since Twitter provides trending topics at the city level and countywide, we can examine the relationship between them. Previous research suggests that there are two mechanisms that drive trend propagation: local diffusion processes and global transmission of trends due to travel hubs~\cite{ferrara2013traveling}. We observed that some trends remained localized; however, the majority of trends achieved nationwide popularity, indicating that the themes spread quite efficiently among the population. Some cities, such as Istanbul, Eskisehir, and Diyarbakir, stand out in terms of their unique trends, while others tend to cluster based on geographic and cultural similarities. We can also track over time how similar these local trends are compared to national trends in Turkey (Fig.\ref{fig:trend-timelines}(right)).

\begin{figure}[t!]
    \centering
    \includegraphics[width=\linewidth]{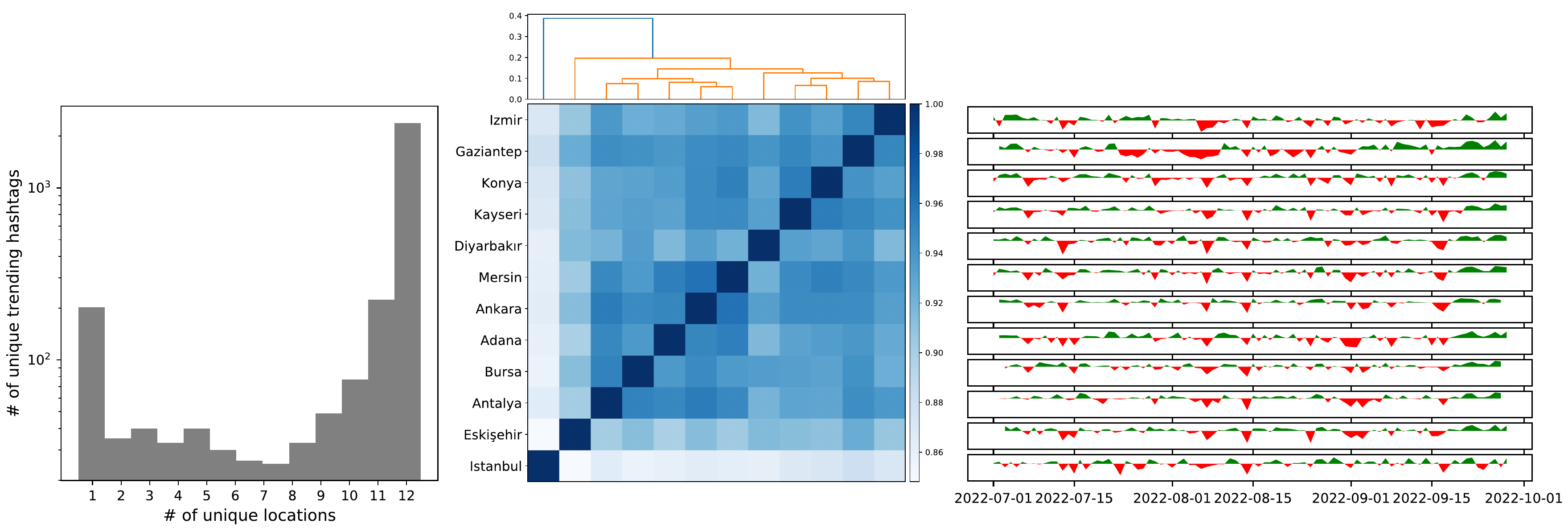}
    \caption{\textbf{Trend topic similarity at the city level.} Some trends are localized meaning appeared only few cities and other reach country-wide popularity (left). Similarities of the trending topics between cities (middle) and their similarity with the Turkey trends (right).}
    \label{fig:trend-similarity}
\end{figure}

\subsection*{Social and information network of Turkish politicians}

Network analysis allows us to observe organizations at the macro and meso levels. These networks can represent a static view of the organization, but can also capture changes over time by using data from different time intervals. Our dataset provides information to create and analyze at least three different networks.

\begin{figure}[t!]
    \centering
    \includegraphics[width=\linewidth]{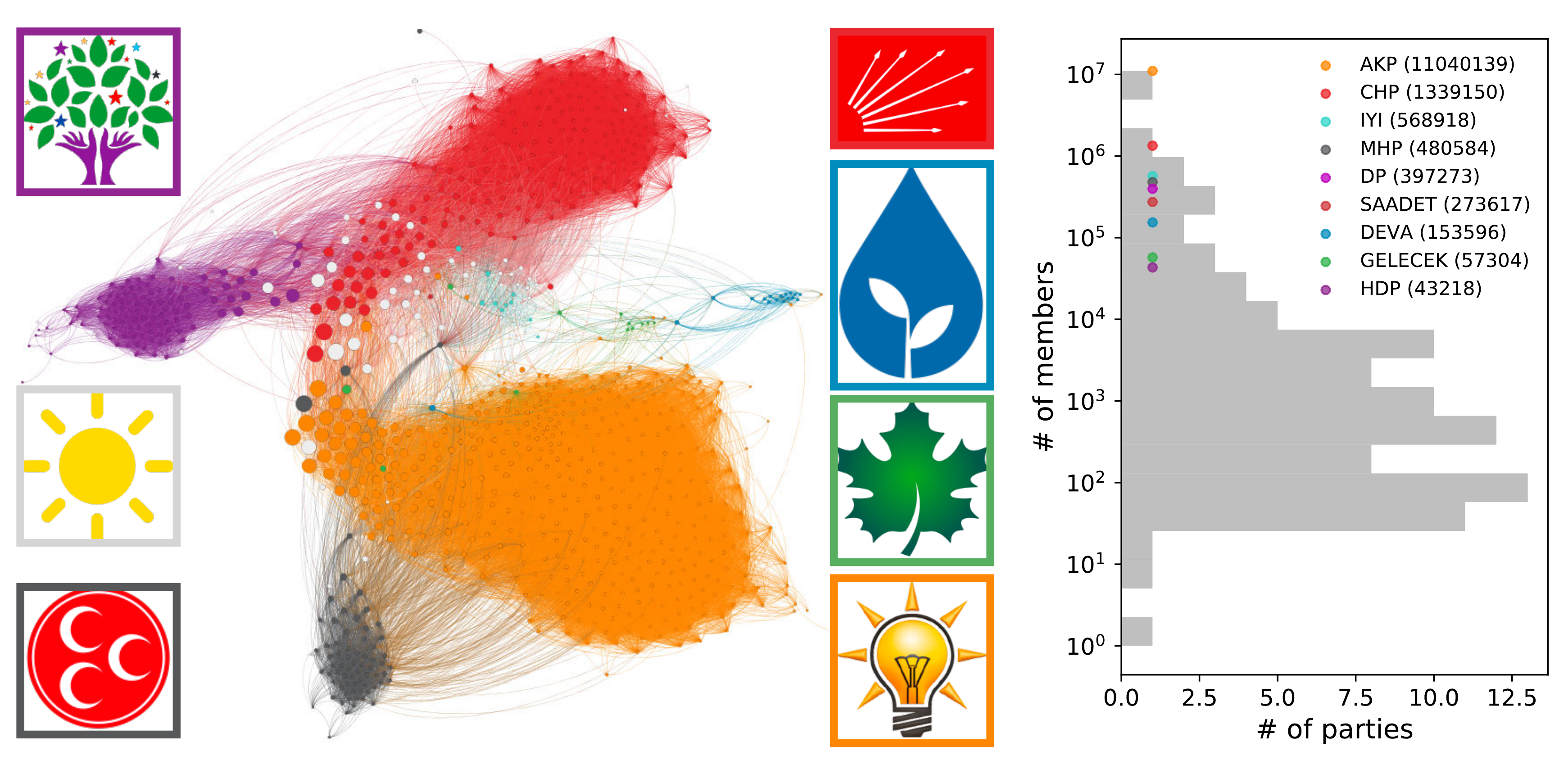}
    \caption{\textbf{Party networks and memberships.} Network nodes present politicians colored to represent their party memberships (left). Registered voter statistics for all Turkish political parties and voter numbers for exemplar parties presented (right).}
    \label{fig:party-network}
\end{figure}

\begin{itemize}
    \item \textbf{Social network}: We regularly collect friends and followers from political accounts, we can build an egocentric network of politicians as well as a similarity network of these political accounts based on their common followers.
    \item \textbf{Information diffusion}: This network captures how information spreads through users by tracking interaction types such as replies, retweets, quotes, and mentions. Nodes can be different tweets or users, and edge weights can represent time delays or frequency of interactions.
    \item \textbf{Hashtag co-occurrence}: Based on the co-occurrence of different memes such as hashtags, URLs, and phrases, we can create a network representation of these memes to show community structures.
\end{itemize}

In Fig.~\ref{fig:party-network} we represent a network of political accounts. In this network, the nodes represent different political figures coloured by their party affiliation. We computed the edge weights as Jaccard similarity between their followers to represent the similarity of their audiences. We applied an additional filtering step to remove edges with low weights by applying a threshold, and used the ForceAtlas2 algorithm as a layout to locate nodes in two-dimensional space. In this network, politicians belonging to the same parties tend to cluster together as partisans share multiple politicians from the same parties. The community organization of this network also provides insight into Turkish politics, as AKP (orange) and MHP (dark grey) represent the People's Alliance, and CHP (red), IYI (light grey), DEVA (blue) from the National Alliance cluster together. The HDP (purple) has distinctly separated from the other two groups, while its politicians share more followers with the politicians from CHP.

\begin{figure}[t!]
    \centering
    \includegraphics[width=\linewidth]{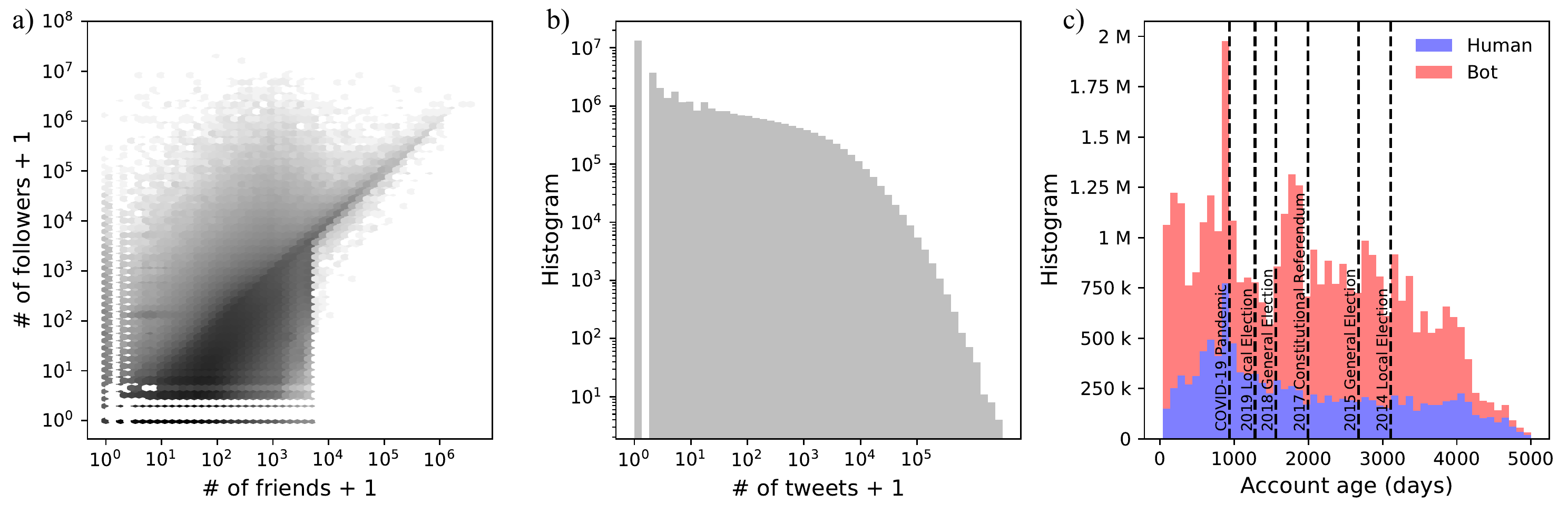}
    \caption{\textbf{Analysis of followers.} Friend and follower statistics (a) and content production measured by tweets and retweets (b) for the followers of politicians presented. We also compared their account creation times and and their use of automation (c).}
    \label{fig:follower-stats}
\end{figure}

Since social networks consist of regular users, their participation in political discourse or engagement with politicians is also important for understanding their representativeness to voters. In Fig.\ref{fig:follower-stats}, we present basic statistics about the following accounts. Most of these accounts have fewer than 1,000 friends and followers. Their productivity follows a heavy-tailed distribution, as most accounts have less than 10 tweets, while very few have more than 100,000 tweets. This discrepancy in content production suggests automated activity~\cite{ferrara2016rise,bastos2019brexit,elmas2022characterizing}. Since our dataset captures these followers at regular intervals, we can examine the deleted network nodes over time.

Analyzing users following political accounts, we observe when they were created in Fig.\ref{fig:follower-stats}(c). We find that more bot accounts than human accounts are created every day since the beginning of 2010. We observe an increase in the creation of automated accounts prior to the 2014 local elections and the 2015 and 2018 general elections. We also note that during the pandemic, the number of human and bot followers increased, and some of these accounts may promote anti-vaccine sentiment and could be repurposed to support certain political ideologies in the upcoming election. Considering our earlier observations of fluctuations in the number of followers and anomalous accounts with extreme content production, it is reasonable to suspect the existence of social bots. 

\subsection*{Automated activities in political networks}

There are several online tools for detecting automated activity, and this is an active area of research given the increasing involvement of automation in political discourse~\cite{ferrara2016rise,cresci2020decade} 
In this work, we use the Botometer system and its light-weight version called \textit{BotometerLite} to evaluate Twitter accounts~\cite{varol2017online,varol2018feature,sayyadiharikandeh2020detection}. This system analyzes user profile information to assess the bot likelihood of an account.

Since social bots can be used to promote politicians, parties, and their agendas, their impact is positive for their campaigns~\cite{vosoughi2018spread,lazer2018science}. However, there are also alternative scenarios in which social bots are used to target politicians, manipulate their engagement rates, and paint a misleading picture of their online presence~\cite{varol2020journalists,varol2018deception}. It is important to consider alternative explanations and collect more evidence to support each claim. Here, we present a brief analysis conducted for 4 major political parties and their politicians.

\begin{figure}[t!]
    \centering
    \includegraphics[width=\linewidth]{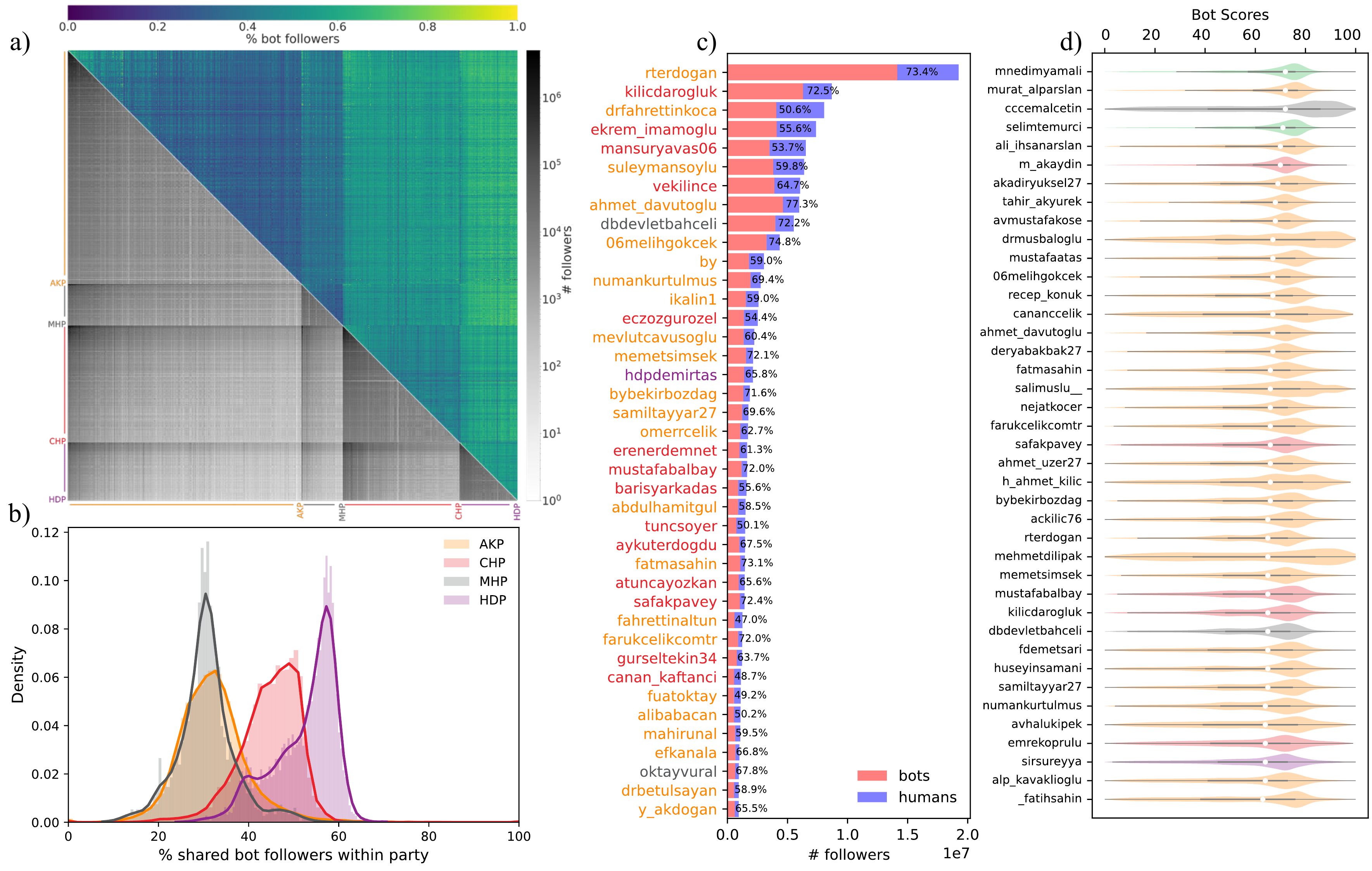}
    \caption{\textbf{Social bot analysis.} Politicians from different parties compared based on number of shared followers and percentage of bots shared (a). Percentage of shared bot followers are also analyzed within parties (b). Individual bot follower statistics investigated by ranking accounts with most followers (c) and highest percentage of bot followers (d).}
    \label{fig:bot-analysis}
\end{figure}

When social bots are studied in the context of politics, the initial question is usually who has the most bot followers. In Fig.\ref{fig:bot-analysis}, we analyzed individual politicians and their aggregate statistics for their parties. Since the prevalence of bot followers can be measured in both exact numbers and percentages, we presented both quantities. In Fig.\ref{fig:bot-analysis} (a), we grouped politicians by their parties and calculated shared number of followers in the lower triangle of the heatmap and plotted the percentage of bot followers among them in the upper triangle of the heatmap. We can see that politicians have more shared followers within their party; however, politicians from CHP and HDP have stronger connections amoung their fellow party members. The percentage of social bots also follows a similar pattern; social connections with CHP and HDP contain more bot accounts. This observation of party connections becomes clearer in Fig.\ref{fig:bot-analysis} (b). It is important to remember that these bots can work for or against these politicians, and this is a research question that we are currently investigating.

When we inspect the most popular individuals (see Fig.\ref{fig:bot-analysis}(c) and the ones with the highest percentage of bot followers (see Fig.\ref{fig:bot-analysis}(d), we observe different sets of names. Popular accounts known to be targeted by social bots to influence their online activities or amplify their engagement metrics and popularity~\cite{varol2020journalists}. We observe a similar result; the popular accounts usually have more than 50\% bot followers. These accounts with high number of followers are mainly from AKP and CHP parties, with only 3 exceptions in the top-40 list. Alternatively, we can also rank politicians by the percentage of bots among their followers. In this figure, we show the distribution of bot scores and the median bot score for these accounts.

An important research work deals with the role of these bot followers. We will investigate information dissemination network using natural language processing tools to determine whether the observed bots work for or against these politicians over the course of the election.

\section*{Conclusion}

In this paper, we introduce the \texttt{Secim2023} dataset and provide a preliminary analysis to highlight potential research questions that can be addressed with this dataset.
For further empirical purposes, researchers can use our data for a variety of research purposes, including network analysis, machine learning applications to predict public sentiment and topics, user demographic data, and election results.

Influence of automated accounts require more in-depth analysis where content analysis, sentiment towards certain parties should be studied with a political science perspective. Our dataset can create opportunities for such interdisciplinary research. 

The dynamic nature of this dataset will support researchers until the election, and we plan to publish regular updates and reports of our findings to the public. Using this dataset, researchers can answer several important research questions about election campaigns, manipulation activities, online trends, etc.
Our team will be using this dataset to work on primarily the following tasks:
\begin{itemize}
    \item Daily activities of social media users to track evolution of topics and the rate of content production various topics.

    \item Track followers of prominent politicians, analyze their social bot followers and report changes in their audience.

    \item Develop tools for early-detection of online manipulation, predicting party affiliations and user demographics.
\end{itemize}

We believe that the \texttt{\#Secim2023} dataset will be a valuable resource for researchers developing natural language processing systems for Turkish and investigating behavior of Turkish speaking social media users using machine learning~\cite{gokcce2014twitter,kaya2013transfer,bulut2017mediatized,polat2014twitter,ccetinkaya2022twitter}.

\textbf{Limitations}: Although we present the most comprehensive and unique dataset to study the 2023 Turkish elections, the dataset may have some limitations. First, our Twitter stream tracks activity across a manually curated list of accounts. We have attempted to capture political figures that represent all of Turkish politics, and we are still collecting recommendations for political figures using a public form published in the data collection section. Second, the Twitter developer agreement limits our ability to share raw data, but researchers can use the Twitter API to rehydrate the original data as long as the tweets are not deleted in the meantime. Finally, Elon Musk's acquisition of Twitter may lead to changes in the platform, the availability of data, and the participation of automated accounts in political discussions~\cite{varol2022should}.

\section*{Author contributions}
All authors contributed data collection from online resources and curation of politician list. AN and OV conducted data analysis. NM and OV wrote the manuscript. OV created study conception and
design.

\section*{Acknowledgments}
We thank the members of the VRL lab, Mert Moral, and the Teyit.org team for their contributions to the initial discussions on the election project. This work was supported in part by the TUBITAK Grant (121C220).

\footnotesize
\bibliographystyle{unsrt}  
\footnotesize
\bibliography{references}

\end{document}